\documentstyle[aaspp4,epsf]{article}

\def\unsetyr{\def\oyear{\relax}\def\cyear{\relax}\def\cyeara{a\relax}\def\cyearb{b\relax}\def\cyearc{c\relax}\def\cyeard{d\relax}\def\cyeare{e\relax}}
\def\setyr{\def\oyear{(}\def\cyear{)}\def\cyeara{a)}\def\cyearb{b)}\def\cyearc{c)}\def\cyeard{d)}\def\cyeare{e)}}

\unsetyr
\def\jcite#1{\setyr\cite{#1}\unsetyr}
\def\rmmat#1{{\hbox{\rm #1}}}
\def\rmscr#1{\rmmat{\scriptsize #1}}
\newcommand{\be}{\begin{equation}}
\newcommand{\ee}{\end{equation}}
\newcommand{\ba}{\begin{eqnarray}}
\newcommand{\ea}{\end{eqnarray}}

\newcommand{\eg}{{\it e.g.~}}
\def\p{\partial}
\def\d{{\rm d}}

\def\pp#1#2{\frac{\p #1}{\p #2}}
\def\eqref#1{Equation~\ref{eq:#1}}
\def\figref#1{Figure~\ref{fig:#1}}

\def\opa{{\tilde \kappa}}

\begin{document}
\newcommand{\bfi}{{\bf B}} \newcommand{\efi}{{\bf E}}
\newcommand{\lel}{{\lambda_e^{\!\!\!\!-}}}
\newcommand{\me}{m_e}
\newcommand{\mcs}{{m_e c^2}}
\title{RCW 103 -- Revisiting a cooling neutron star}
\author{Jeremy S. Heyl\altaffilmark{1}}
\authoremail{jsheyl@tapir.caltech.edu}
\author{Lars Hernquist}
\authoremail{lars@ucolick.org}
\affil{Lick Observatory,
University of California, Santa Cruz, California 95064, USA}
\altaffiltext{1}{Current address: Theoretical Astrophysics, mail code 130-33,
California Institute of Technology, Pasadena, CA 91125}

\begin{abstract}

Recent observations of the compact source embedded within the supernova
remnant RCW 103 rekindle interest in the origin of this object's
emission.   We contrast several models in which neutron-star cooling
powers RCW 103.  Specifically, either the presence of an accreted
envelope or a sufficiently intense magnetic field can account for the
X-ray emission from this object.
\end{abstract}


\section{Introduction}

Soon after the X-ray source 1E~161348-5055 was first detected by the
{\it Einstein} observatory (\cite{Tuoh80}) near the center of the
supernova remnant (SNR) RCW 103, \jcite{Tuoh83} proposed that this
source is an isolated neutron star emitting thermal radiation.
Optical and radio observations have failed to identify a counterpart
(\cite{Tuoh80}; \cite{Tuoh83}; \cite{Dick96}; \cite{Kasp96}),
bolstering the interpretation of this source as an isolated neutron
star.  Subsequent X-ray observations with {\it Einstein} and {\it
ROSAT} have not all confirmed the initial detection (\cite{Tuoh80};
\cite{Beck93}).

Using recent observations of 1E~161348-5055 with the {\it ASCA}
observatory and archival data from {\it ROSAT}, \jcite{Gott97} verify
the existence of this source and refocus attention on the interpretation
of its emission.  After subtracting a model for the emission of the
surrounding SNR, \jcite{Gott97} find that the point source spectrum is
well described by a blackbody having a characteristic temperature $kT =
0.6$ keV and a flux of $6 \times 10^{-12}$ erg s$^{-1}$ cm$^{-2}$.
Estimates of the distance to RCW 103 vary from 3.3 kpc (\cite{Casw75})
to 6.6 kpc (\cite{Leib83}).  Combining these values yields an estimated
luminosity of $8 d_{3.3}^2 \times 10^{33}$ erg s$^{-1}$ and an effective
emitting area of $7 d_{3.3}^2 \times 10^{10}$ cm$^2$ where $d_{3.3}$ is
the ratio of the true distance to the X-ray source to 3.3 kpc.  This is
less than a percent of the total surface area of a neutron star.  So,
unless the emission originates from a tiny hotspot, a blackbody cannot
account for the emergent spectrum.  \jcite{Gott97} also find no periodic
variation in the flux greater than 13 \% of the mean count rate.
Variation of this order or larger would be expected from a rotating
neutron star emitting from a small portion of its surface unless the hot
spot coincides with the rotation axis, the object's period is outside
the range explored, or gravitational defocusing smooths the periodic
signal. 

Several models for this object have been proposed since its discovery.
\jcite{Gott97} argue that the object is unlikely to be a cooling
neutron star, a plerion, or a neutron star with an ordinary companion.
The dismissal of these models prompted \jcite{Popo97} to argue that
the emission from 1E~161348-5055 is powered by accretion onto a
neutron star in a binary with another compact object.  Unless the
magnetic field of the neutron star is exceptionally weak ($B <
10^8$~G), it will channel the accreted material onto the polar caps
producing hotspots and variability.  For an apparently young
object to have such a weak field, he argues that the neutron star is
{\it not} coeval with the remnant, but that the remnant resulted from
the supernova of the binary companion.

In this {\it Letter}, we revisit models of 1E~161348-5055 which
account for its emission through neutron star cooling.  In the first,
a neutron star cooling through an accreted envelope naturally results
in a spectrum which greatly departs from a blackbody.  The second
model, an ultramagnetized cooling neutron star, results in anisotropic
emission from a hotspot with a spectrum which qualitatively resembles
a blackbody.

\section{Analytic Models}

The insulating envelope of a neutron star may be modeled analytically
if the magnetic field is sufficiently strong or weak.  We use the
models of \jcite{Hern84b} to describe the heat transport through an
unmagnetized hydrogen envelope.  The ultramagnetized models of
\jcite{Heyl97analns} describe the relationship between core
temperature and transmitted flux for ultramagnetized envelopes
($B\gtrsim 10^{15}$ G).  For such intense magnetic fields, these
analytic models agree well with fully numerical calculations
(\cite{Heyl98numens}).  Both sets of calculations adopt a plane
parallel approximation to solve the thermal structure equation in the
envelope and assume that the passage from the non-degenerate to the
degenerate regime is abrupt.

The equations are solved from the surface toward the core using a
zero flux boundary condition (\cite{Schw58}).  The core temperature is a
function of the combination $F/g_s$ and $\psi$ (in the magnetized case).
Here, $F$ is the transmitted heat flux, $g_s$ is the surface gravity, and
$\psi$ is the angle between the radial and field directions, and all of
these values are taken to be in the frame of the neutron star surface.
In the zero field limit, the models are independent of angle and in the
ultramagnetized limit $F/g_s \propto \cos^2 \psi$ for a fixed core
temperature. 

\subsection{An Accreted Envelope}

As a baseline model, we consider a neutron star that has accreted
sufficient unprocessed material since its birth to have a light-element
envelope.  We consider a hydrogen layer which extends all the way down to
a density $\rho \sim 10^{10}$ g cm$^{-3}$.  The total mass of accreted
hydrogen would be $\sim 10^{-8} {\rm M}_\odot$ (\cite{Heyl97kes}).

To obtain the core temperature as a function of transmitted flux, we
recalculate the models of \jcite{Hern84b} with the atomic number ($Z$) and
weight ($A$) equal to unity.  Since \jcite{Hern84b} examine iron
envelopes for which $Z+1 \approx Z$, we have to make several alterations to
their results to account for the pressure contributed by the ions in the non-
degenerate regime.

Specifically, if the conductivity has a power-law
form,
\be
\kappa = \kappa_0 \frac{T^\beta}{\rho^{\alpha}},
\ee
then the relation of temperature to density in the non-degenerate
regime follows a solution such that the conductivity is a constant
\be
\kappa = \frac{\alpha + \beta}{\alpha} \frac{F}{g_s} \frac{(Z+1) k}{A m_p}.
\ee
The relationship between $T$ and $\rho$ takes the form,
\be
T = \left ( \frac{\alpha + \beta}{\alpha} \frac{F}{g_s} \frac{(Z+1) k}{A
m_p}\frac{1}{\kappa_0} \right )^{1/\beta} \rho^{\alpha/\beta}.
\label{eq:TrhoND}
\ee
We assume that free-free scattering dominates the opacity through the
non-degenerate portion of the envelope; consequently,
\be
\alpha = 2, \beta = \frac{13}{2}, \kappa_0 = \frac{16 \sigma}{3} m_u
\frac{196.5}{24.59} \frac{A^2}{Z^3}
\frac{\rmmat{g}}{\rmmat{cm}^5\rmmat{K}^{7/2}}.
\ee

Since electrons dominate the pressure in the degenerate regime, we use
equations (2.22) through (2.29) of \jcite{Hern84b} without alteration to
determine the core temperature given the temperature at the onset of
degeneracy calculated using \eqref{TrhoND}.  The more detailed
unmagnetized models of \jcite{Pote97} show that neutron stars with
partially accreted envelopes exhibit cooling evolution between that of
objects with fully accreted envelopes and standard cooling scenarios.

\subsection{An Ultramagnetized Iron Envelope}

We use the calculations for ultramagnetized iron envelopes described in 
\jcite{Heyl97analns}.  As in the unmagnetized case, we assume that free-free 
scattering dominates the opacity in the non-degenerate regime.
However, we apply an anisotropy factor to account for the effect of the
magnetic field on the scattering rates.  We use the results of
\jcite{Pavl76} and \jcite{Sila80} to estimate this effect.  

In the degenerate regime, we proceed analytically to a density where the first
Landau level fills.  Below this density, the relationship between 
chemical potential and density is analytically invertible in the fully
degenerate limit; consequently, for an iron envelope with $B=10^{16}$
G, we can analytically integrate the structure equations up to a
density of $1.5 \times 10^{10}$ g cm$^{-3}$ using the conductivities
for the liquid and solid phases presented by \jcite{Hern84a}.
For the magnetized envelopes the emission is anisotropic 
along the surface.  The average flux over the entire surface is a
factor of 0.4765 times its peak value at the magnetic poles (\cite{Heyl97analns}).

\begin{figure}
\plotone{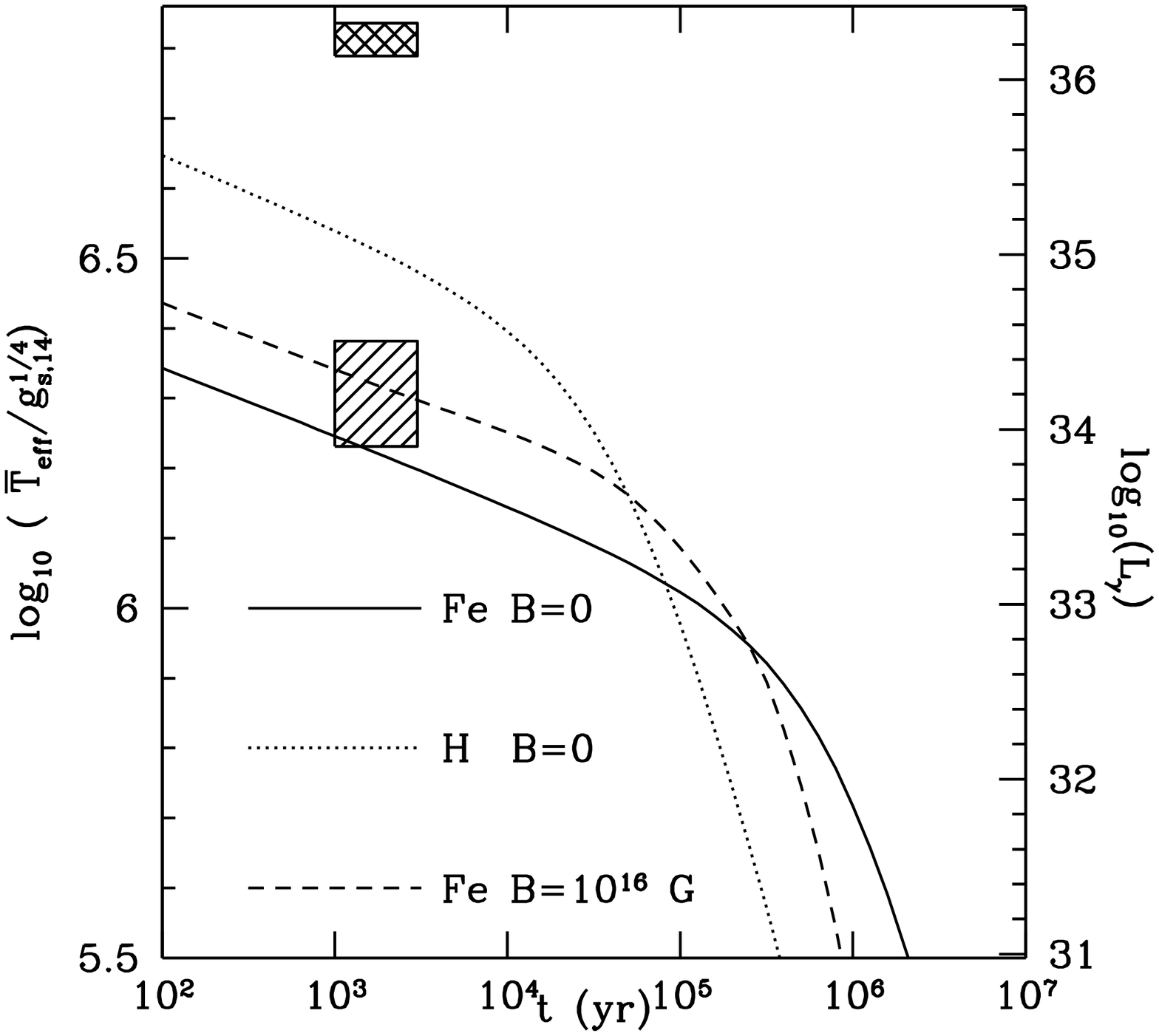}
\caption[dummy]{
The cooling evolution calculated analytically
for iron and hydrogen envelopes.  The results for unmagnetized iron
envelopes are taken from \jcite{Hern84b}.  The upper cross-hatched
region shows the fitted blackbody temperature of
1E~161348-5055 and the acceptable range in age from 1,000 to
3,000 years.  The lower shaded region depicts the luminosity of the
object estimated from observations.
}

\label{fig:coolanal}
\end{figure}

With each of these models we determine the relationship between
$T_\rmscr{eff}$ and $T_c$ for $10^5 \rmmat{K} < T_\rmscr{eff} < 10^7
\rmmat{K}$ and calculate the cooling curves as described in
\jcite{Heyl97magnetar}.  \figref{coolanal} traces the cooling
evolution for both ultramagnetized and unmagnetized iron envelopes,
and unmagnetized hydrogen envelopes.  We find that an intense magnetic
field increases the emitted flux from an neutron star at a given time
during the neutrino-cooling epoch by an amount sufficient to account
for the observed luminosity of the point source in RCW~103 of $\sim
10^{34}$ erg s$^{-1}$, if the neutron star is approximately 1,000
years old.  This age estimate is consistent with the observations of
the remnant (\cite{Tuoh79}; \cite{Nuge84}; \cite{Cart97}).  However,
the effective temperature, $kT=0.2$ keV, falls short of the observed
characteristic blackbody temperature of $0.6$ keV.

The presence of a fully accreted envelope dramatically increases the
effective temperature for a given core temperature.  For a
1,000-year-old neutron star the effective temperature $kT=0.3$ keV is only
a factor of two below that observed.  The total luminosity is
$\sim 10^{35}$ erg s$^{-1}$, greatly exceeding that observed.  

Due to gravitational fractionation, a neutron star which has accreted any
hydrogen will have a hydrogen atmosphere.  The presence of a hydrogen
atmosphere shifts the emission blueward from that of a blackbody.

\subsection{Analytic power-law atmosphere}

In the LTE structure and NLTE spectrum formation limit of studying an
atmosphere, it is straightforward to derive the spectrum for power-law
conductivities.  In general the conductivity and its associated
opacities are given by
\be
\kappa = \kappa_0 \frac{T^\beta}{\rho^\alpha}, \opa =
\frac{16 \sigma}{3} \frac{T^3}{\rho \kappa} \rmmat{~and~}
\opa_E = f(\gamma) \opa \left ( \frac{E}{k T} \right )^{-\gamma}
\ee
where $\opa_E$ is the opacity as function of photon energy and
$f(\gamma)$ is obtained by calculating the Rosseland mean
\be
f(\gamma) = \int_0^\infty \pp{B}{T} \left ( \frac{E}{k T} \right
)^{\gamma} \d E \Biggr / \int_0^\infty \pp{B}{T} \d E 
          = \frac{15}{4 \pi^4} \int_0^\infty
	\frac{x^4 e^x x^\gamma}{\left(e^x-1\right )^2} \d x
\ee
where $B(E)$ is the Planck function for the intensity of blackbody
radiation.
 
We know that along a solution to LTE structure equations
$\d T/\d z = F / \kappa$ where both $F$ the flux and $\kappa$ are constant.
The optical depth at a given energy is given by
\be
\tau(E,T) = \int_0^z \opa_E \rho \d z = \int_0^T \opa_E
\frac{\kappa}{F} \rho \d T' = \int_0^T \frac{16}{3} \frac{T'^{3+\gamma}}{T_\rmscr{eff}^4}
f(\gamma) T_E^{-\gamma} \d T' = f(\gamma) \frac{1}{4+\gamma} \frac{16}{3} \left (\frac{T}{T_\rmscr{eff}}
\right )^4 \left ( \frac{T_E}{T} \right )^{-\gamma}
\ee
where we have changed variables, assuming the radiative zero solution, and
substituted for $\opa_E$ and $F=\sigma T_\rmscr{eff}^4$.  We define
$T_E=E/k$. 

In the limit of a blackbody atmosphere, we have $\gamma=0$,
$f(\gamma)=1$, and $T=T_\rmscr{eff}$ where the spectrum forms,
therefore we take $\tau=4/3$
to determine the temperature at which photons of a given energy have
their effective photosphere.  After some rearrangement we obtain,
\be
\frac{T_\rmscr{atm}}{T_\rmscr{eff}} = \left ( f(\gamma) \frac{4}{4+\gamma} \right
)^{-1/(4+\gamma)}
\left ( \frac{T_E}{T_\rmscr{eff}} \right )^{\gamma/(4+\gamma)}.
\ee

\begin{figure}
\plottwo{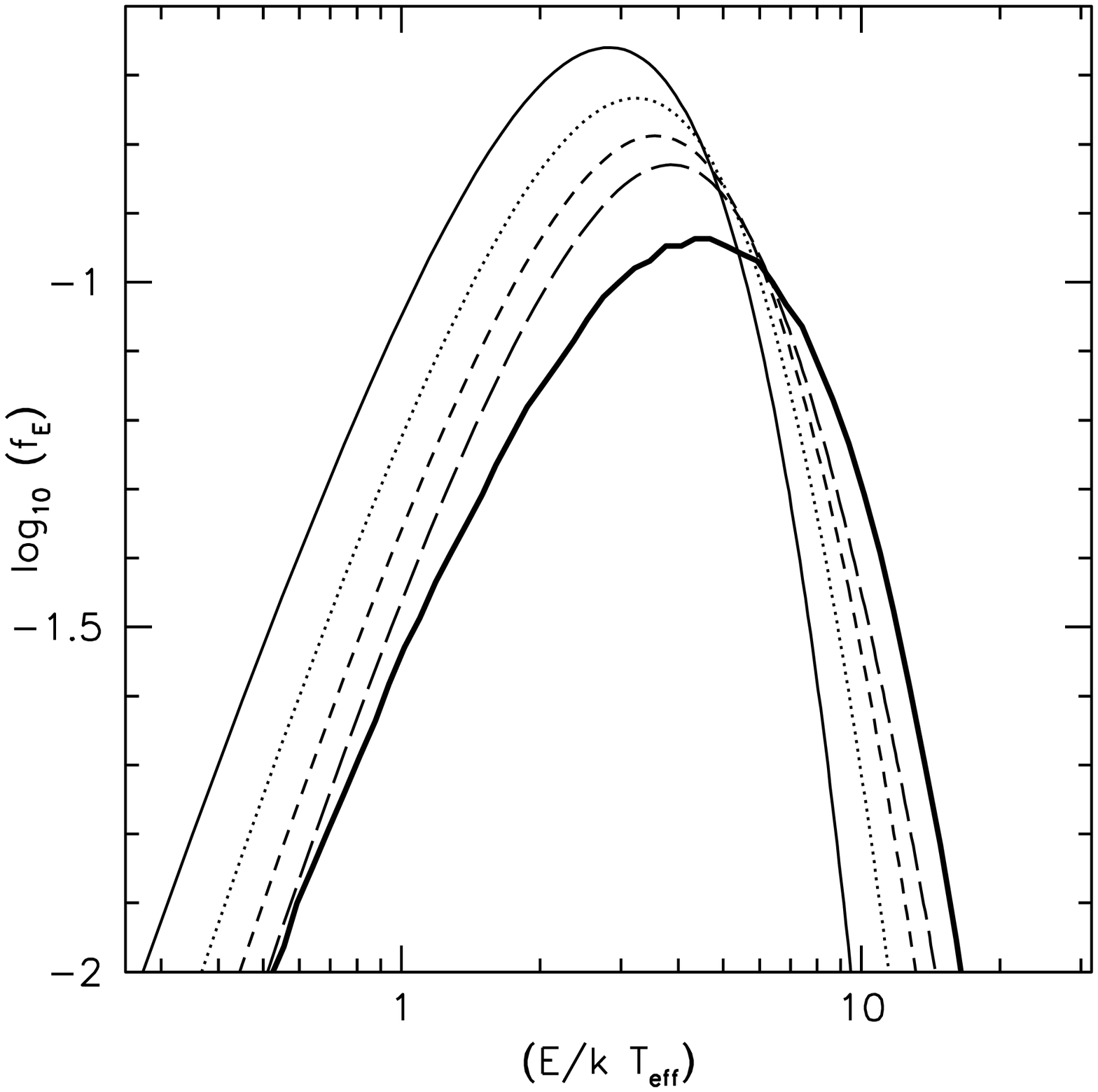}{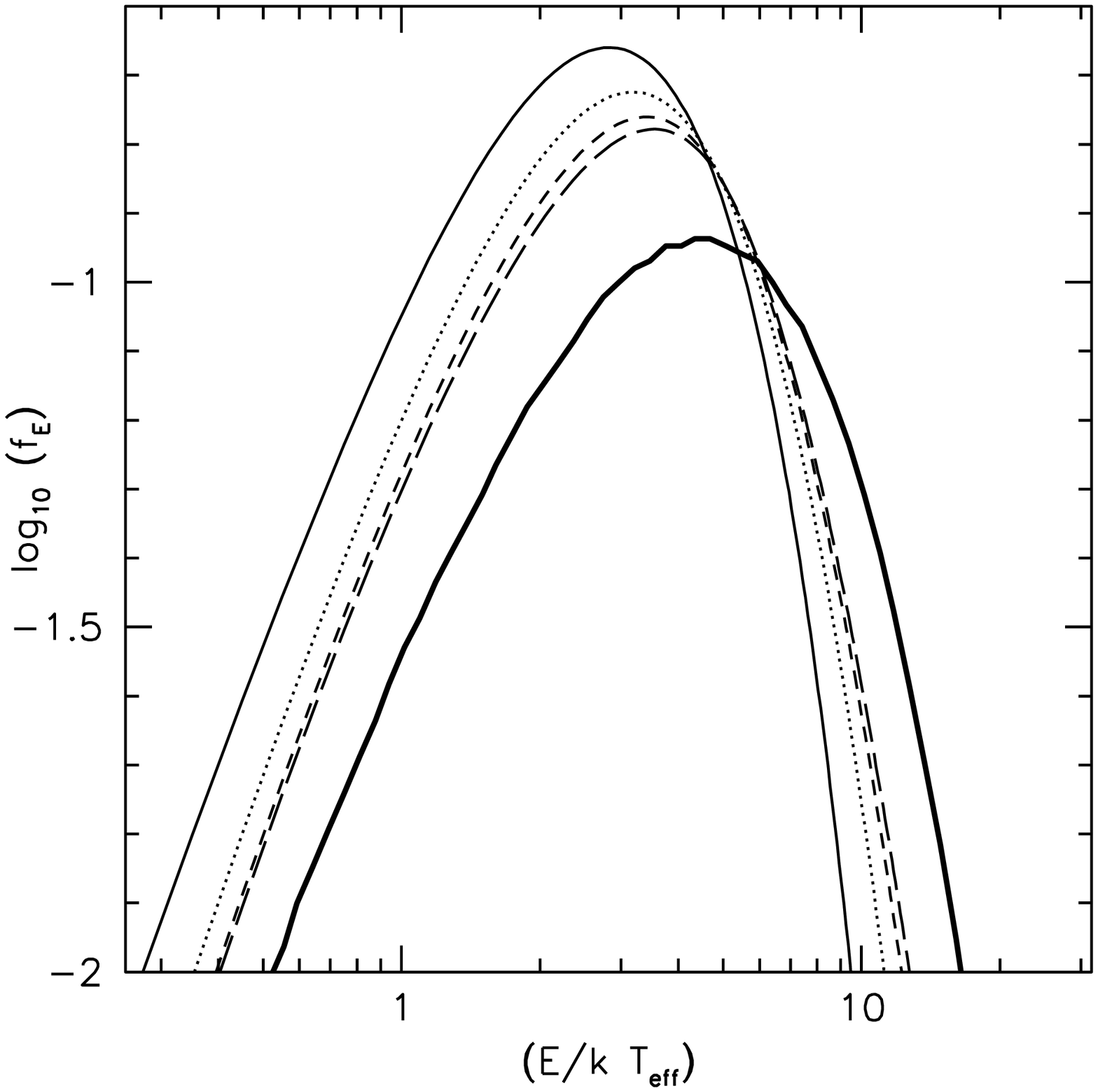}
\caption[dummy]{
Emergent spectra from the power-law atmospheres.  The left panel
shows the direct calculation, and the right depicts the iterated
results.  The solid line traces a blackbody spectrum, the dotted line
is for a $\gamma=1$ opacity, the short-dashed line is $\gamma=2$, and
the long-dashed line is $\gamma=3$.  The heavy solid line shows
results for a hydrogen atmosphere with
$T_\rmscr{eff}=10^{6.5}$~K and $g_s=2.43 \times 10^{14}$~cm~s$^{-2}$
from \jcite{Zavl96}.
}
\label{fig:atmo}
\end{figure}

To calculate the spectrum, we make the following three assumptions
\begin{enumerate}
\item
The Rosseland mean opacity is unaffected even though the spectrum
diverges from a blackbody through the atmosphere,
\item
If the optical depth for a photon of a given energy to escape to
infinity is less than $4/3$, it is completely free, and
\item
If the optical depth is greater than or equal to $4/3$, the photon
is drawn from a blackbody distribution at the appropriate temperature.
\end{enumerate}
At each layer of the atmosphere, flux conservation is imposed by
scaling the blackbody contribution to the flux such that the total
flux is equal to $\sigma T_\rmscr{eff}^4$.  With this prescription, it
is straightforward to derive the emergent spectra as depicted in left
panel of \figref{atmo}.  Even in this simple model, we find that an
energy dependent opacity can shift the emergent spectra blueward of a
blackbody.  We find for $\gamma=3$, appropriate for free-free
scattering (\cite{Kipp90}) that the peak is shifted blueward by nearly
40 \%, which would result in an underestimate of the emission region
by a factor of about four.

To understand the dependence of the models on the first assumption, we
can calculate the function $f(\gamma)$ using the emergent spectrum
rather than a blackbody.  By iterating from five to ten times,
we find the appropriate values of $f(\gamma)$ to within one part per
thousand.  These iterated spectra are depicted in the right panel of
\figref{atmo}.  For all values of $\gamma$ the blueward shift is less
pronounced than in the direct prescription.  For example for
$\gamma=3$, the peak is shifted blueward by 25 \%, which would result
in an underestimate of the emitting area by a factor of 2.5.  A
self-consistent power-law atmosphere would require a recalculation of
the mean opacity at each depth of the atmosphere.  Within the model,
according to assumptions (2) and (3), the spectrum at a given depth
consists partly of a blackbody and partly of the emergent spectrum;
therefore, we expect that a spectrum calculated self-consistently
throughout the atmosphere would fall between these two limiting cases.

More detailed modeling of hydrogen atmospheres supports our
conclusions.  As shown in \figref{atmo}, this effect is
even more pronounced when one examines the detailed calculations of
\jcite{Zavl96}.   They find that if one fits a blackbody to the
emergent hydrogen spectrum one will overestimate the effective
temperature by an even larger factor, $\sim 4.2$, than found in our
models.  Consequently, an ultramagnetized neutron star with an
iron envelope and a hydrogen atmosphere can account for both the
observed flux and characteristic temperature of 1E~161348-5055.

\section{Conclusions}

We find that a young neutron star cooling through a strongly magnetized
or a partially accreted envelope can account for the observed emission
from 1E~161348-5055.  The detailed models of \jcite{Pote97} support the
conclusions that we have found analytically in this {\it Letter}.
Furthermore, the estimates of the emitting area of 1E~161348-5055
support the conclusion that the spectrum from this object is
significantly harder than a blackbody and possibly results from
emission through a hydrogen atmosphere.

The recent discovery by \jcite{Tori98} of a 69-ms X-ray pulsar
(J161730-505505) in the vicinity of SNR RCW 103 complicates the
evaluation of the possible models.  It has a spin-down age of $8.1
\times 10^3$~yr, several times larger than that of the remnant.
\jcite{Tori98} examine the possibility that the X-ray pulsar is
associated with the remnant, and find that a kick velocity of $1300
d_{3.3} t_{8.1}^{-1}$ km s$^{-1}$ is required to explain its current
position relative to the center of the supernova remnant.  Such a
large supernova kick velocity is uncommon but has been observed for
other pulsars (\cite{Lyne94}).  However, when this object is compared
with other similar rotation-powered, plerionic pulsars, it is a factor
of ten underluminous.  We would argue with \jcite{Gott97} that this
source is a heavily absorbed background object, possibly a
rotation-powered, plerionic pulsar as \jcite{Tori98} suggest but located
at a distance $\sim 10$~kpc.  A reanalysis of the spectrum studied by
\jcite{Tori98} may be able to determine the interstellar column
density to J161730-505505 and verify its status as a background
source.

\jcite{Gott97} failed to find flux variations at a level of 13 \% over a
wide range of periods.  Although the total flux from a magnetized
neutron star may not vary at this level because of gravitational
defocussing (\eg \cite{Heyl97analns}), the magnetic field causes the
atmospheric emission to be highly anisotropic (\cite{Shib95};
\cite{Raja97}), so its apparent lack of variability may indicate that it
is only weakly magnetized ($B \sim 10^{11}$~G) or simply that the
geometry is not conducive to large flux variations.  Observations of this
object with AXAF should be able to distinguish between these models by
determining the spectral shape of 1E~161348-5055.

We argue that the X-ray source in the supernova remnant RCW 103 is
simply the natural end product of stellar evolution through a supernova:
an isolated, cooling neutron star.  

\acknowledgements

We would like to thank the referee G. Chabrier for useful suggestions
that improved the original manuscript.  The work was supported in part
by a National Science Foundation Graduate Research Fellowship and Cal
Space grant CS-12-97.

%\cleardoublepage

\end{document}